# Improving Unsupervised Stain-To-Stain Translation using Self-Supervision and Meta-Learning

Nassim Bouteldja[a,b,e], Barbara M. Klinkhammer[b], Tarek Schlaich[a], Peter Boor[b,c,*], Dorit Merhof[a,d,*]

[a] Institute of Imaging and Computer Vision, RWTH Aachen University, Aachen, Germany
[b] Institute of Pathology, RWTH Aachen University Hospital, Aachen, Germany
[c] Department of Nephrology and Immunology, RWTH Aachen University Hospital, Aachen, Germany
[d] Fraunhofer Institute for Digital Medicine MEVIS, Bremen, Germany
[e] Corresponding author: nbouteldja@ukaachen.de, University Hospital Aachen, Pauwelsstr. 16, 52074 Aachen, Germany

* Authors contributed equally

# Abstract

**Background:** In digital pathology, many image analysis tasks are challenged by the need for large and time-consuming manual data annotations to cope with various sources of variability in the image domain. Unsupervised domain adaptation based on image-to-image translation is gaining importance in this field by addressing variabilities without the manual overhead. Here, we tackle the variation of different histological stains by unsupervised stain-to-stain translation to enable a stain-independent applicability of a deep learning segmentation model.

**Methods:** We use CycleGANs for stain-to-stain translation in kidney histopathology, and propose two novel approaches to improve translational effectivity. First, we integrate a prior segmentation network into the CycleGAN for a self-supervised, application-oriented optimization of translation through semantic guidance, and second, we incorporate extra channels to the translation output to implicitly separate artificial meta-information otherwise encoded for tackling underdetermined reconstructions.

**Results:** The latter showed partially superior performances to the unmodified CycleGAN, but the former performed best in all stains providing instance-level Dice scores ranging between 78% and 92% for most kidney structures, such as glomeruli, tubules, and veins. However, CycleGANs showed only limited performance in the translation of other structures, e.g. arteries. Our study also found somewhat lower performance for all structures in all stains when compared to segmentation in the original stain.

**Conclusions:** Our study suggests that with current unsupervised technologies, it seems unlikely to produce "generally" applicable simulated stains.

# 1. Introduction

Histological analysis represents the current gold standard for tissue examination in research and diagnostics.[1] The field of digital pathology is steadily growing, particularly since it enables automated and reproducible high-throughput analysis of highly resolved tissue data. Due to the widespread dissemination of digital whole slide scanners, large amounts of histological data can be obtained in clinical routine and preclinical research. This process also includes several degrees of variability, e.g. in staining protocols, dye compositions, cutting thicknesses, pathological alterations, and scanner characteristics. This poses a substantial challenge for image analysis, since tackling all sources of variation with manual, time-consuming efforts is not feasible. Thus, the field of unsupervised domain adaptation based on image-to-image translation has gained tremendous popularity in recent years.[2] This field comprises methods that convert between different image domains, e.g. horse and zebra images, by transferring the image style, e.g. translating a horse to a zebra or vice versa. Since this is performed unsupervisedly, i.e. without the need for any data annotations, it enables a significant reduction of manual overhead for image analysis. Such approaches are frequently applied in digital pathology in four main areas, (1) stain normalization,[3-8] i.e. the task of reducing color variations within a particular stain, (2) stain translation,[9-12] i.e. the field of compensating variation across different staining protocols, including (3) the conversion between histology and fluorescence[13, 14] or (4) between simulated masks and the image domain.[12, 15-17] Here, the cycle-consistent generative adversarial network (CycleGAN) is most frequently employed, i.e. an approach for training convolutional neural networks (CNNs) for image-to-image translation between two domains, as it represents the state-of-the-art technique for unsupervised domain adaptation and demonstrated the feasibility of producing realistic image translations.[18]

In this work, we focus on stain translation approaches using CycleGANs to enable image analysis on differently stained histological data in an uninformed fashion[19], i.e. no ground-truth labels are required for the target stain. In experimental and clinical histopathology, numerous dyes are used to stain specimens resulting in differently colored and textured tissue (Fig. 1). Artificial Intelligence (AI) applications in digital pathology have predominantly been trained on single stains and cannot intrinsically cope with inter-stain variation. Strengthened by the projected decrease in pathologist workforce[20], exhaustively performing sufficient data annotations for each particular stain is not feasible and also prevents analysis on newly developed stains. Thus, there is a great need for yielding stain independence in CNNs[21,22]. This bears the potential to include diagnostically-relevant molecular information into the analysis, e.g. as is required for the diagnosis of kidney allograft rejection[21], and could thus leverage new possibilities for the broad implementation in digital pathology.

1.1. Medical Background

In histopathology, specimens are always stained first using general-purpose stains such as hematoxylin-eosin (HE) or periodic acid-Schiff (PAS) for an initial high-contrast visualization of various tissue structures. Often, immunohistochemical stains (IHC) are additionally employed for the detection of specific target proteins facilitating in-depth analysis. In this work, we focus our study on kidney pathology, in which PAS represents the most suitable and most widely used staining. Histological analysis of major kidney structures, particularly glomeruli, tubules, and interstitium (Fig. 1) is an essential part of histopathological diagnostics, with various diseases affecting various compartments in different ways. Whereas the tubules occupy most of the kidney tissue (of up to 75%[23]), glomeruli cover only a few percent of tissue, similarly to arteries. Morphometric analysis of these structures can provide valuable information on the pathomechanisms of renal disease. To perform and facilitate such analyses, the renal structures need to be segmented.

1.2. Related work

In total, three groups have recently reported the feasibility of segmenting major renal structures using CNNs,[23-25] more precisely U-nets[26] with modified architectures. Jayapandian et al.[24] trained their U-nets on single structures, however failed to separate touching instances from one another. In contrast, Hermsen et al.[25] and Bouteldja et al.[23] added an artificial border class around structures to enable instance segmentation. All three groups trained their networks on single stains resulting in insufficient generalization capabilities across other stains.

Gupta et al.[27] have tackled this issue by aligning arbitrarily stained slides to the analyzable stain using a cubic B-spline-based registration approach. However, whole-slide images often contain individual artifacts that likely limit registration performance. In addition, the approach expects consecutive slides, which is an expensive requirement often not met. To prevent this, Gadermayr et al.[9] introduced stain-to-stain translation using CycleGANs and could also report slightly better segmentation performance of glomerular tufts.

Several further domain adaptation approaches have been applied in digital pathology[2]. For stain normalization, Shaban et al.[4] and de Bel et al.[5] used CycleGANs to transfer single-stained data between different scanners (Aperio and Hamamatsu) as well as centers, respectively. Salehi et al.[3] and Cho et al.[8] converted HE stained images into grayscale and employed a conditional GAN-like framework[28] to revert the conversion. Whereas the generator in Salehi et al.[3] learned the mapping to a normalized HE representation by tackling this underdetermined problem, the generator in Cho et al.[8] was trained to map various stain styles to a specific one to learn its color distribution. The authors further penalized differences in features between the input image and its mapping extracted from a tumor classification network to preserve relevant features for classification on normalized images.[8] Nonetheless, stain normalization approaches aim to address color variations only within a particular stain.

Regarding fluorescence translation, Burlingame et al.[14] used the pix2pix framework[29] to translate between immunofluorescent stains and HE, and Rivenson et al.[13] employed adversarial training to map autofluorescence images to HE. In both works, the authors reported promising translation results and pointed out the potential of fluorescence translation to omit the need for clinical multiplexing and histological staining procedures. But in contrast to IHC staining, fluorescent imaging allows for the generation of corresponding image pairs by additional registration of both domains, thus enabling the use of such supervised techniques.

Furthermore, Gadermayr et al.[15] and Bug et al.[16] simulated mask images showing rather simple elliptical structures and transferred them to the image domain using CycleGANs to enable an unsupervised segmentation. Both groups improved translation by incorporating low-level features such as nuclei simulations into the mask domain. However, such approaches are limited to the analysis of simple structures that can be modeled mathematically.

Regarding the application of stain translation, de Haan et al.[30] used generative adversarial networks including CycleGANs to translate HE stained kidney biopsies into three simulated stains, i.e. Masson's trichrome, PAS, and Silver, that were altogether considered for improving the preliminary diagnosis of non-neoplastic diseases. Further, Gadermayr et al.[12] and Lo et al.[10] trained a model for the segmentation of glomeruli on a single stain (PAS and HE, respectively) and then enabled its application on various other stains by translating them into the single analyzable stain using CycleGANs. The former[12] additionally showed that this direction of translation worked far better than translating the single annotated stain to the others for training stain-specific segmentation models on simulated data. They also illustrated the importance of translating into an easy-to-segment stain such as PAS for segmentation. In our study, we will follow up on these findings. However, both works[10,12] only demonstrated the feasibility of stain translation for the analysis of a single structure, and further paved the way for investigations of integrating the segmentation model into the translator for its improvement.

### 1.3. Our contributions to stain translation

Our objective is to make the supervised PAS segmentation network from Bouteldja et al.[23] applicable to various other stains without any further manual annotaton effort by employing CycleGANs for unsupervised stain-to-stain translation following Gadermayr et al.[12] However, we examine its feasibility for the segmentation of various renal structures including tubules, glomeruli, glomerular tufts, arteries, arterial lumina, and veins (Fig. 1). We further propose two novel approaches to improve translational efficiency. First, we integrate the pre-trained segmentation model[23] into the translation network in a self-supervised manner. The aim is to support a proper translation of those structures through semantic guidance and to motivate mappings closer to the learned distribution of the segmentation model for improved applicability. Second, we propose to tackle the limitation in cycle-consistency-based training of assuming bijective mappings, by providing extra channels to both domains that can be used to separate artificial and interfering meta-information from the translation. Next to reporting qualitative and quantitative improvements, we also compare our baseline with the U-GAT-IT model[31] that appeared to have strongly outperformed the CycleGAN.

## 2. Methods

In our application scenario, we assume a pre-trained (segmentation) model $S : P \to L$ that allows for the analysis of a specific stain $P$. We aim at making it applicable to an arbitrary stain $A$ by using a CycleGAN to translate between both stains.

2.1. CycleGANs

The CycleGAN[18] is a type of generative adversarial network that is widely applied for unsupervised style transfer. It performs image-to-image translation between two image domains (here: stains $P$ and $A$) using unpaired data and consists of two generators $G_{P \to A}, G_{A \to P}$ and two discriminators $D_P, D_A$. The networks are trained by optimizing the following three losses: The adversarial loss

$$L_{adv} = \mathbb{E}_{x \sim p_P(x)} \left[ \log(D_P(x)) + \log\left(1 - D_A(G_{P \to A}(x))\right) \right]$$
$$+ \mathbb{E}_{y \sim p_A(y)} \left[ \log(D_A(y)) + \log\left(1 - D_P(G_{A \to P}(y))\right) \right]$$

makes the generators produce realistic (simulated) images $G_{P \to A}(x), G_{A \to P}(y)$ with respect to the target domain, while the discriminators aim to differentiate between those translations and real images $x \in P, y \in A$. The cycle consistency loss

$$L_{cyc} = \mathbb{E}_{x \sim p_P(x)} \left[ \|G_{A \to P}(G_{P \to A}(x)) - x\|_1 \right] + \mathbb{E}_{y \sim p_A(y)} \left[ \|G_{P \to A}(G_{A \to P}(y)) - y\|_1 \right]$$

represents the core idea of CycleGANs as it forces the generators to reconstruct their input when being subsequently forwarded through both of them. In this way, each generator learns the inverse of the other's mapping. Spatial consistency between input and translation is implicitly encouraged due to its simplicity in learning. Finally, the identity loss

$$L_{idt} = \mathbb{E}_{x \sim p_P(x)}[\|G_{A \to P}(x) - x\|_1] + \mathbb{E}_{y \sim p_A(y)}[\|G_{P \to A}(y) - y\|_1]$$

incentivizes both generators to forward images from the target domain unchanged and is shown to improve color preservation as well as training stability.[32]

2.2. Self-supervision

We integrate the assumed segmentation network $S$, which outputs label probability maps for samples from the analyzable stain $P$, into the CycleGAN as depicted in Fig. 2 (exemplarily with $P$ and $A$ representing the PAS and CD31 stain). During training iterations, only images from $P$ as well as their reconstructions and identity mappings are further propagated through the segmentor $S$. Since we can assume the segmentation results of real samples to represent the ground truth pretty well, we use them as targets for self-supervision by penalizing their discrepancies to the respective predictions using the following segmentation loss $L_{seg}$:

$$L_{seg} = L_{seg,cyc} + L_{seg,idt}$$
$$= \mathbb{E}_{x \sim p_P(x)} \left[ \left\| S\left(G_{A \to P}(G_{P \to A}(x))\right) - S(x) \right\|_1 + \left\| S(G_{A \to P}(x)) - S(x) \right\|_1 \right]$$

The motivation for enforcing equal segmentation predictions on reconstructions and identities of $P$ is twofold. First, when considering that the segmentor $S$ has only been trained on real samples from $P$, its applicability to simulated images (using $G_{A \to P}$) potentially originating from a different probability distribution might be impeded. Despite visual similarities between simulated and real samples, the translator could still encode unnatural information and (noise) patterns into the simulated images due to the imperfection of adversarial training. Since this type of information is unfamiliar to $S$, it might harm performance. To prevent this, our proposed segmentation loss $L_{seg}$ encourages the generator $G_{A \to P}$ to project its translations into the learned source distribution of $S$, hence translated images are better analyzable by $S$. Second, optimizing $L_{seg}$ also helps leverage semantic features to better learn the concepts and translational correspondences of the classes. The predicted segmentation targets semantically guide the generators to properly translate the different class structures by bringing attention to their mappings. This application-oriented guidance could particularly tackle confusion in the translation of underrepresented classes.

Using loss-specific weights, the overall loss function $L$ can now be formulated as:

$$L = \lambda_{adv} L_{adv} + \lambda_{cyc} L_{cyc} + \lambda_{idt} L_{idt} + \lambda_{seg} L_{seg} .$$

2.3. Meta-learning

As stated in Section 2.1., CycleGANs consist of two generators that each aim to learn the other's inverse as triggered by the cycle consistency loss. Hence, the underlying assumption and limitation of this framework is that both mappings between the domains $G_{P \to A}$, $G_{A \to P}$ represent bijections. However, this does not apply for the underdetermined translation between general-purpose and IHC stains. The latter can provide molecular information especially in pathological structures or stain-specific arbitrary artifacts that cannot be inferred from the general-purpose stain. The generators tackle this challenge of mapping from an information-rich to an information-poor domain by encoding source domain-specific information, typically in a visually imperceptible manner, into the translations to enable a subsequent, well-defined reconstruction.[33] On the example of the underdetermined zebra-to-horse translation studied by the CycleGAN authors,[18] the first generator maps a zebra to a horse image and additionally encodes the information about the stripes into the translated horse, so that the second generator can then use it to enable a precise reconstruction of the same zebra. Regarding our application scenario, such unnatural encodings of structure into translated images are unfamiliar to the subsequently applied segmentation model and would most likely decrease its performance. Even in case of visually imperceptible information, the harm could be extensive as shown by adversarial worst-case examples.[34]

To tackle this, we propose adding three extra feature channels (analogous to image size) to both the input ($M_1$) and output ($M_2$) of each generator, with the input zero-padded by three extra channels (Fig. 3). The output now consists of the usual three-channel translation that is propagated through the respective discriminator, but also of the three additional channels $M_2$ that can be used to store useful meta-information from the input for reconstruction. The subsequent generator then back-propagates them both for a well-defined reconstruction of the input. Overall, this provides the opportunity for the generators to implicitly decouple artificial meta-information from the translations to make them more realistic and thus better usable by a subsequent model. It is noteworthy that the generators could now simply copy the input into the extra channels $M_2$ and then copy it back for perfect reconstruction. However, the fact that the generators also must manage image translation without any additional information (i.e., on zero-padded inputs), prevents this undesirable side effect.

2.4. Evaluation

Our objective is to enable and improve the applicability of an already existing segmentation model $S$ to arbitrary unknown stains. For $S$, we employ the U-Net-like model from Bouteldja et al.[23] that has been trained on kidney tissue stained in PAS and reported high performance for the instance segmentation of various renal structures. Likewise, we also use instance-level Dice scores to measure the segmentation accuracy of $S$ on translated PAS images from various stains. For a set of test images $t \in T$ and their respective binary instance predictions $p_{t,i}$ and ground-truths $g_{t,j}$ indexed by $i = 0, ..., n_{p_t}$ and $j = 0, ..., n_{g_t}$, the instance-level Dice score (IDSC) is computed for each class as follows:

$$IDSC = \frac{1}{\sum_{t \in T} n_{p_t} + n_{g_t}} \sum_{t \in T} \left( \sum_{i}^{n_{p_t}} DSC(p_{t,i}, g_{t,*}) + \sum_{j}^{n_{g_t}} DSC(g_{t,j}, p_{t,*}) \right)$$

Here, $n_{p_t}$ and $n_{g_t}$ represent the numbers of prediction and ground-truth instances for image $t$, and $g_{t,*}$ stands for the ground-truth instance with maximal overlap to prediction instance $p_{t,i}$ (0 if false positive), vice versa for $p_{t,*}$. Analogous to Dice scores (DSC), the IDSC ranges between 0 (no single overlap in all test images) and 1 (perfect overlaps). By averaging the equally weighted Dice scores for prediction and ground-truth instances across the whole test set, the IDSC measures the mean area overlap per instance.

The segmentation accuracies on translated (simulated) PAS images directly infer the translational performance in terms of feature preservation of the predicted structures. Thus, we expect the reported segmentation performance of $S$ on real PAS images[23] to represent the upper accuracy limit for our translation approach providing simulated data. We also use t-tests for the comparison between the unmodified CycleGAN and our proposed models by pairwise comparison of the underlying Dice score distributions of each class.

2.5. Data

Paraffin-embedded kidney tissues from mice were cut into 1-2 μm thick sections that were digitized by the NanoZoomer C9600-12 whole-slide scanner (Hamamatsu Corporation, Bridgewater, New Jersey) with a 20x objective lens after staining. The employed stains included PAS as commonly used for overview staining in kidneys, cluster of differentiation (CD31) highlighting endothelial cells, alpha-smooth muscle actin (aSMA) as a marker for smooth muscle cells, collagen III (Col3) highlighting fibrosis, and finally neutrophil gelatinase-associated lipocalin (NGAL) as a marker of tubular cell injury. In total, our in-house data set comprised 85 whole-slide images (WSIs) divided into 53 PAS, 8 CD31, 7 aSMA, 10 Col3, and 7 NGAL WSIs. Regarding our application scenario, we trained models for stain translation between PAS and IHC (CD31, aSMA, Col3, NGAL), respectively, to translate the latter into the PAS domain for subsequent segmentation by $S$. We used all PAS slides and randomly chosen 5 CD31, 5 aSMA, 7 Col3, and 5 NGAL WSIs for training and the remaining slides (3 CD31, 2 aSMA, 3 Col3, 2 NGAL) for evaluation of the stain translators. Our data preprocessing pipeline started with the gray-scale conversion and Otsu's thresholding for automated tissue detection in WSIs. We then performed image tessellation to extract patches of size 216 µm x 216 µm and resampled them into images of 640x640 pixel resolution, overall complying with the input requirements of $S$. In total, 35233 PAS, 3104 CD31, 2969 aSMA, 5533 Col3 and 3857 NGAL patches were extracted for training. For testing, we manually annotated 20 patches in each IHC evaluation slide in QuPath,[35] a widely used open-source software in digital pathology. The annotation procedure was performed as described and defined in Bouteldja et al.[23] The resulting 200 annotated patches (60 CD31, 40 aSMA, 60 Col3, 40 NGAL) were considered the ground truth and compared with the segmentation predictions on their corresponding simulated PAS translations to finally quantify the performance of $S$ on the IHC stains.

2.6. Experimental setting

In our experiments, we trained CycleGANs for stain translation with and without our proposed modifications of incorporating the prior segmentation model as well as multi-channels into training, respectively. Since training was conducted on 640x640-pixel images, we slightly adapted the employed CycleGAN architecture from Gadermayr et al.[12] We incremented the depth of both U-Net-based generators to seven and the depth of the PatchGAN discriminators to four to enlarge their receptive fields accordingly. The networks were trained for 300,000 iterations using a batch size of three and RAdam[36] as optimizer. After 150,000 iterations, the initial learning rate of $10^{-4}$ started to linearly decrease to zero until the last iteration. In accordance with Gadermayr et al.,[12] we also used equally weighted loss terms ($\lambda_{adv} = \lambda_{cyc} = \lambda_{idt} = \lambda_{seg} = 1$) and employed standard data augmentation (flipping, 90° rotation, gamma correction).

In addition, we trained an Unsupervised Generative Attentional Network with Adaptive Layer-Instance Normalization for Image-to-Image Translation (U-GAT-IT) model[31] as a baseline for the unmodified CycleGAN due to its reported promising superiority. The U-GAT-IT extends the CycleGAN by integrating auxiliary domain classifiers and class activation maps[37] into generators and discriminators to focus on discriminative and thus relevant image regions for translation. However, for the vanilla U-GAT-IT, we experienced severe training instability issues for the generators showing few regular loss spikes and providing unrealistic translations. Thus, we searched for its optimal configuration based on training behavior. We replaced the ResNet-based generators with the same U-Net architectures as utilized in our CycleGAN models and used a network depth of three and four for the two employed scales of PatchGAN discriminators, respectively. The replacement of ResNet- with U-Net-based generators fixed the instability issues.

The training settings were kept the same in all experiments. All technical details about the utilized segmentation network are further described in Bouteldja et al.[23] In summary, a U-Net-like model was trained on a large set of heavily augmented PAS-stained kidney tissue (primarily from mice) for instance segmentation of multiple renal structures including tubules, glomeruli, arteries, arterial lumina, and veins (Fig. 1).

All experiments were implemented in PyTorch and were conducted on an NVIDIA A100 GPU (requiring about 7, 11, and 20 GB of VRAM for the CycleGAN, its incorporation of the segmentation network, and the U-GAT-IT, respectively). We made our code publicly available at (https://github.com/NBouteldja/KidneyStainTranslation).

## 3. Results

Quantitative segmentation performance showed relatively high instance-level accuracies in all classes and stains, except for arteries and arterial lumina that were predicted with considerably worse performance, especially in Col3 (Table 1). In comparison, worse predictions were obtained for the segmentation of tubules in aSMA and glomerular structures in Col3 and aSMA. Among all evaluated models, the CycleGAN variant that solely incorporates the segmentation network showed the highest mean performance across all classes in each stain and provided high instance-level Dice scores ranging between 78% and 92% for non-arterial structures. The incorporation of extra channels demonstrated improvements in CD31 and aSMA over the unmodified CycleGAN, but in contrast decreased performance of the best performing model (CycleGAN w/ SegNet) in all stains. Interestingly, the unmodified CycleGAN baseline proved to be superior to the U-GAT-IT model in all stains. The performance quantifications showed high standard deviations and only a few statistically significant differences. Besides, the employed segmentation network performed substantially better for all structures when being applied on real PAS images.[23]

Qualitative translation and segmentation results using the best performing model (CycleGAN w/ SegNet) are depicted in Fig. 4 for all stains, which show different characteristics in color and texture. Regarding row one to four, the simulated PAS translations of all IHC stains appear highly realistic and detailed, and thus resulted in predictions close to the ground-truth. In all stains, especially in those providing homogenous color transitions between touching tubular instances (e.g., NGAL), the translators managed to generate contrastive borders around the tubules (representing the tubular basement membrane). This enabled the segmentation network to separate those instances from one another. By overlaying IHC inputs onto their translations (column two), we observed a high degree of spatial consistency that is the prerequisite for the transferability of segmentation results to the original IHC image. The last two rows show the reasons for the stain-related, bad-performing trends mentioned above. In Col3, most arteries were translated in such a manner that they had unnatural, tubular-like substructures (e.g. tubular cytoplasm) in their muscle layer (row five, left arrow). These were responsible, on the one hand, for incorrect tubule predictions (row five, right arrow), and on the other hand, for the missed identification of the arterial wall resulting in the confusion of arterial lumina with veins. Here, the translator seemed to fail at learning the translational class concepts of arteries and tubules and confused them with each other. In addition, we observed tubular translations showing unnatural patterns of cytoplasm (i.e. its gray filling) in aSMA (row six, left arrows) that also confused the segmentation model and made it miss those.

Further qualitative segmentation and translation results of all evaluated models are comparatively shown in Fig. 5. Although the unmodified CycleGAN provided a realistic translation of that individual glomerulus on a visual basis, the segmentation model could not make use of it at all. Despite the visual similarities, the translations of all other proposed CycleGAN variants were actually predictable by the segmentation network. The U-GAT-IT model provided a translation that was hardly distinguishable from that ineffective one, however still showed partial signs of glomerular predictability. Regarding the variants incorporating extra channels for their intended purpose of learning meta-information (rows three and four), we identified a few artifacts in their PAS translations (red arrows) likely leading to false predictions (last row, right arrow). Contrary to our expectations, their extra channels appeared to encode various structural information and provided slight global artifacts in a grid-like pattern.

## 4. Discussion

In this work, we enabled and improved the applicability of a previously existing segmentation network trained on PAS to different IHC stains by using novel CycleGAN approaches for unsupervised stain translation. Our aim was to improve state-of-the-art methods for stain independence to open new possibilities for low-cost computational analyses in digital pathology, such as an automated large-scale morphometric analysis of immunostainings.

Although both modifications yielded performance improvements on their own, solely incorporating the segmentation network $S$ represented the best performing model in all stains. Nevertheless, all models showed severe limitations in a proper translation of arteries in Col3 as they have been confused with tubular patterns ultimately preventing their identification. Even the implicit guidance by $S$ for a predictable arterial translation was not sufficient to solve these shortcomings, indicating that a higher degree of supervision is required here. In addition, compared to the reported segmentation performance of $S$ on real PAS samples,[23] our results were inferior in all classes. This demonstrates the presence of feature differences in real and simulated samples that are relevant for segmentation, and opens future research perspectives for further improvements in stain translation. This finding as well as the outcome of confused arterial translations with tubular structures in Col3 also provide the answer to a question raised in Tschuchnig et al.,[2] that a translation from an IHC stain (e.g. Col3) to a general-purpose stain (e.g. PAS) is only partially capable of showing similar features as the real target stain. Considering the incorporation of extra channels, we observed improved segmentation performances only in CD31 and aSMA, which most probably indicates that the benefit of separating meta-information from translations depends on the degree of underdetermination in the reconstruction of the specific stain. Hence, we forecast that in translational applications showing more underdetermined relations (e.g. horse-to-zebra mappings[18]), this approach could leverage its potentials and provide more promising results. Besides, qualitative translation results showed a few interfering artifacts and the encoded information in the extra channels has not met our expectation of encoding only structures with ambiguous reconstructions (e.g., certain Col3-positive areas). A possible explanation for the wide range of structural information that has been encoded, instead, is that the translator might have additionally tackled the variability of color intensities by storing those inside the extra channels and outputting normalized PAS translations to some extent.

Our best performing model (CycleGAN w/ SegNet) achieved a performance of 89% instance-level Dice scores for glomerular tuft segmentation in CD31, which appeared to be superior to the B-spline registration-based approach[27] that yielded Dice scores of 83% for the same task. It also does not require PAS-stained consecutive slides.

Quantitative performances confirmed that the translators effectively made use of the immunohistochemical highlighting of structures of interest. E.g., aSMA highlighted the muscle cells in arteries that have been predicted considerably better than in all other stains, and CD31 marked endothelial cells that are components of glomerular tufts and closely connected to arterial lumina, and thus led to an improved prediction of those structures. This shows that attention is implicitly brought to the highlighted structures that facilitates learning of their translational correspondences. However, although aSMA sensitively identifies arteries, their predictions have still not achieved the comparative performances of $S$ on real PAS images.[23] One plausible reason for this may be the non-specificity of the stain resulting in other structures also being positively highlighted to some extent, which makes learning of translational correspondences more difficult.

The qualitative translation results demonstrated realistic simulated PAS translations from various IHC stains. This raises the question of how similar those translations are to real samples with a view of visual and especially sub-visual features, and whether the similarity is sufficient for the applications of arbitrary PAS networks without loss of performance. In this regard, effectively measuring the quality of GAN-based synthesized images is still an open field of research. Although visual evaluations by physicians previously showed the indistinguishability of real and simulated samples in similar stain translation tasks,[10] our qualitative results demonstrated that subtle, potentially imperceptible changes can, however, affect the predictability of structures. The high standard deviations of Dice score distributions in all classes confirm this finding of translating structures either in a predictable or non-predictable way. Gadermayr et al.[12] translated Col3 to PAS using a vanilla CycleGAN and showed the predictability of glomeruli on realistically looking translated images by a prior segmentation model. This led to their conclusion that CycleGAN-based image-to-image translation can be performed highly effectively to convert between different stains. According to our results, a simulated PAS translation from Col3 was not feasible for the segmentation of arteries as they were partly confused with tubular structures. Taken together, predicting which tissue structures are ideally translated by which model, and what kind of information might be lost, seems not to be feasible. Therefore, to study the similarities between real and simulated samples, metric learning techniques between image distributions may be a promising approach.

## 5. Conclusions

In this work, we investigated CycleGANs for unsupervised stain-to-stain translation in digital pathology to facilitate a stain-independent segmentation of various renal structures, and further proposed two novel approaches to improve translational effectivity. The model solely incorporating the segmentation network performed best in all stains yielding instance-level Dice scores ranging between 78% and 92% for glomerular structures, tubules, and veins. This suggests that translation can be boosted in an application-oriented manner that will require adapted models for each specific task. However, the translation of arteries revealed the limitations of CycleGAN-based stain translation as they were partly confused with tubular structures and thus poorly predicted. Further core findings were that subtle changes to realistically translated structures suffice to toggle their predictability, and that translation is the limiting factor here compared to segmentation. Our study suggests that with current unsupervised technologies, it seems very challenging to produce "generally" applicable simulated stains. In future work, we intend to facilitate a proper translation of arteries by improving the degree of guidance using semi-supervised learning.

## Authors' contribution

NB, DM and PB planned and oversaw the study. NB planned and conducted experiments, NB and TS performed annotations. NB performed statistical analyses. BMK and PB aquired data. NB wrote the first draft of the manuscript and arranged figures. PB and DM critically reviewed the manuscript and figures. All authors read and approved the final version of the article.

## Acknowledgments

This study was funded by the German Research Foundation (DFG) (project numbers: 233509121; 445703531; 322900939; 454024652), the "Exploratory Research Space (ERS)" of RWTH Aachen University (project number OPSF585), the German Federal Ministries of Education and Research (BMBF: STOPFSGS-01GM1901A), Health (DEEP LIVER, ZMVI1-2520DAT111) and Economic Affairs and Energy (EMPAIA) and the European Research Council (ERC) under the European Union's Horizon 2020 research and innovation program (grant agreement No 101001791).

## Disclosure

The authors declare that there is nothing to disclose.

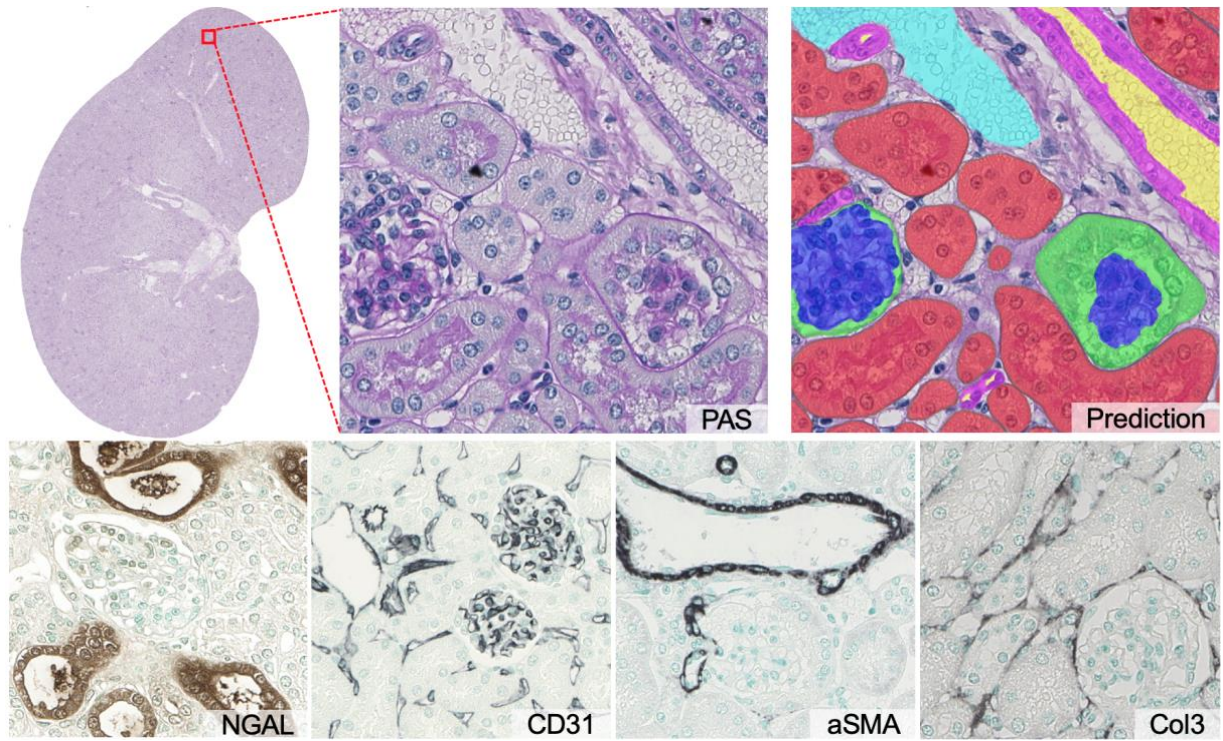

**Figure 1: Overview of analyzed stains and renal structures.**

Our study focuses on histopathological kidney tissue stained in periodic acid-Schiff (PAS), neutrophil gelatinase-associated lipocalin (NGAL), cluster-of-differentiation 31 (CD31), alpha-smooth muscle action (aSMA), or collagen III (Col3), altogether providing distinct differences in color and texture. We are specifically interested in the instance segmentation of various renal structures including tubules (colored red in the upper right image), glomerular tufts (blue), full glomeruli (green + blue), veins (cyan), arterial lumina (yellow), and arteries (magenta + yellow) in all stains. The preliminary segmentation model[23] performed this task with high accuracies solely on the PAS stain (prediction depicted in the upper right image), which is why this work aims to make it applicable to the other stains.

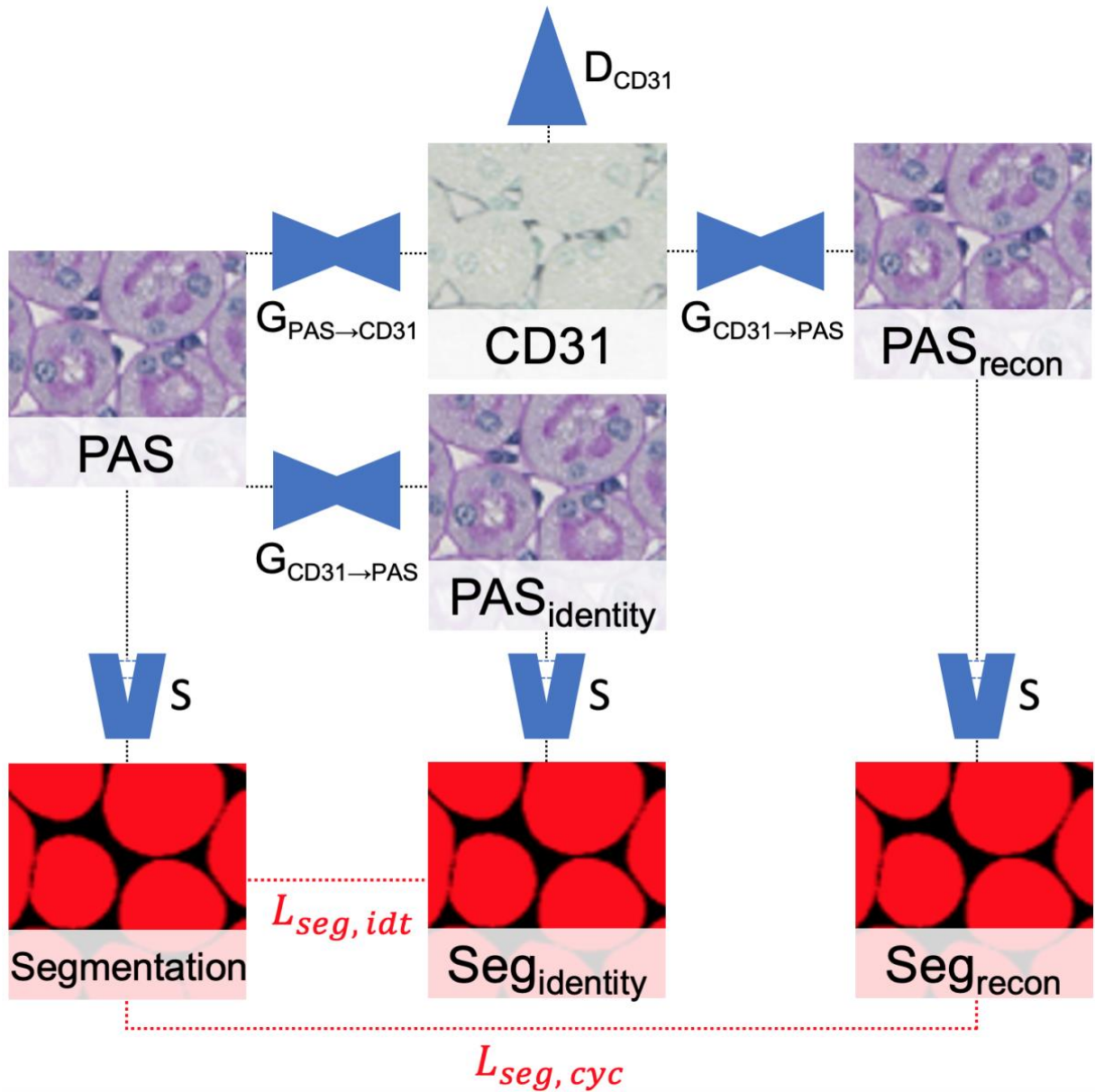

**Figure 2: Prior segmentation model for semantic guidance.**

The outline of the proposed integration of a prior segmentation model $S$ into the CycleGAN on the example translation between PAS and CD31 is shown. During training, $S$ is integrated in a one-sided fashion: Only for PAS inputs, segmentation predictions are performed using $S$ and treated as ground-truth for the predictions on their reconstructions as well as identity mappings. Their discrepancies are penalized using the $\ell_1$-losses $L_{seg,cyc}, L_{seg,idt}$, respectively.

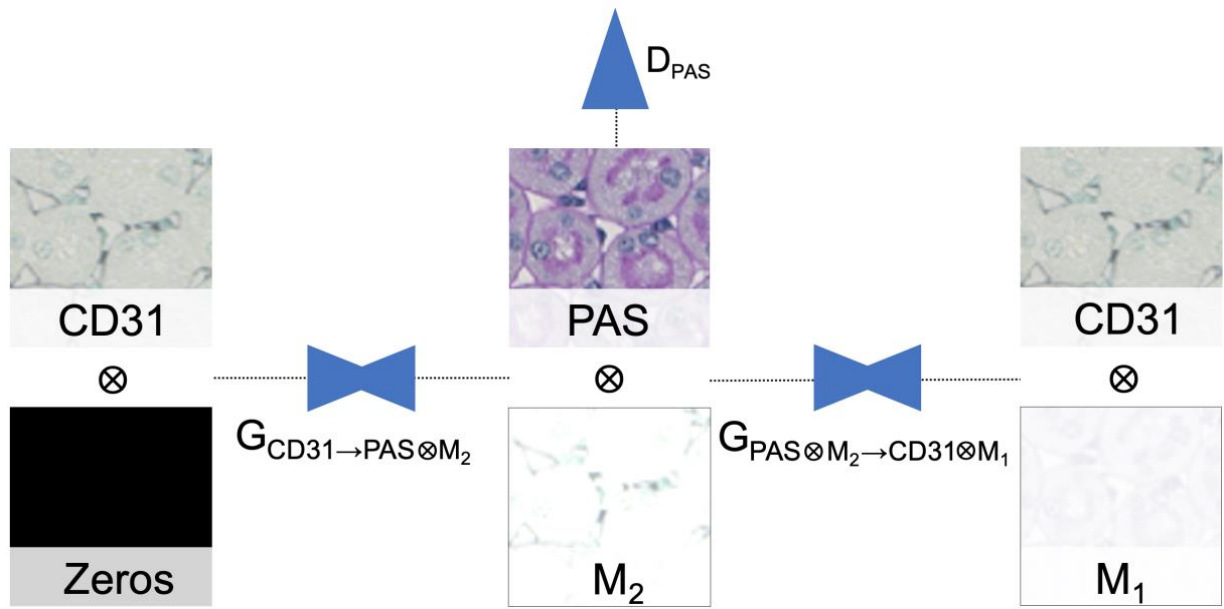

**Figure 3: Extra channels for meta-learning.**

The proposed incorporation of extra channels into the CycleGAN is exemplified for one of the two translational directions. The input is zero-padded by three channels and the output now includes the translation as well as three additional channels $M_2$ that can be used to implicitly learn meta-information. Both are then translated back for input reconstruction (ignoring in turn its extra channels $M_1$). Here, $\otimes$ represents the concatenation operation.

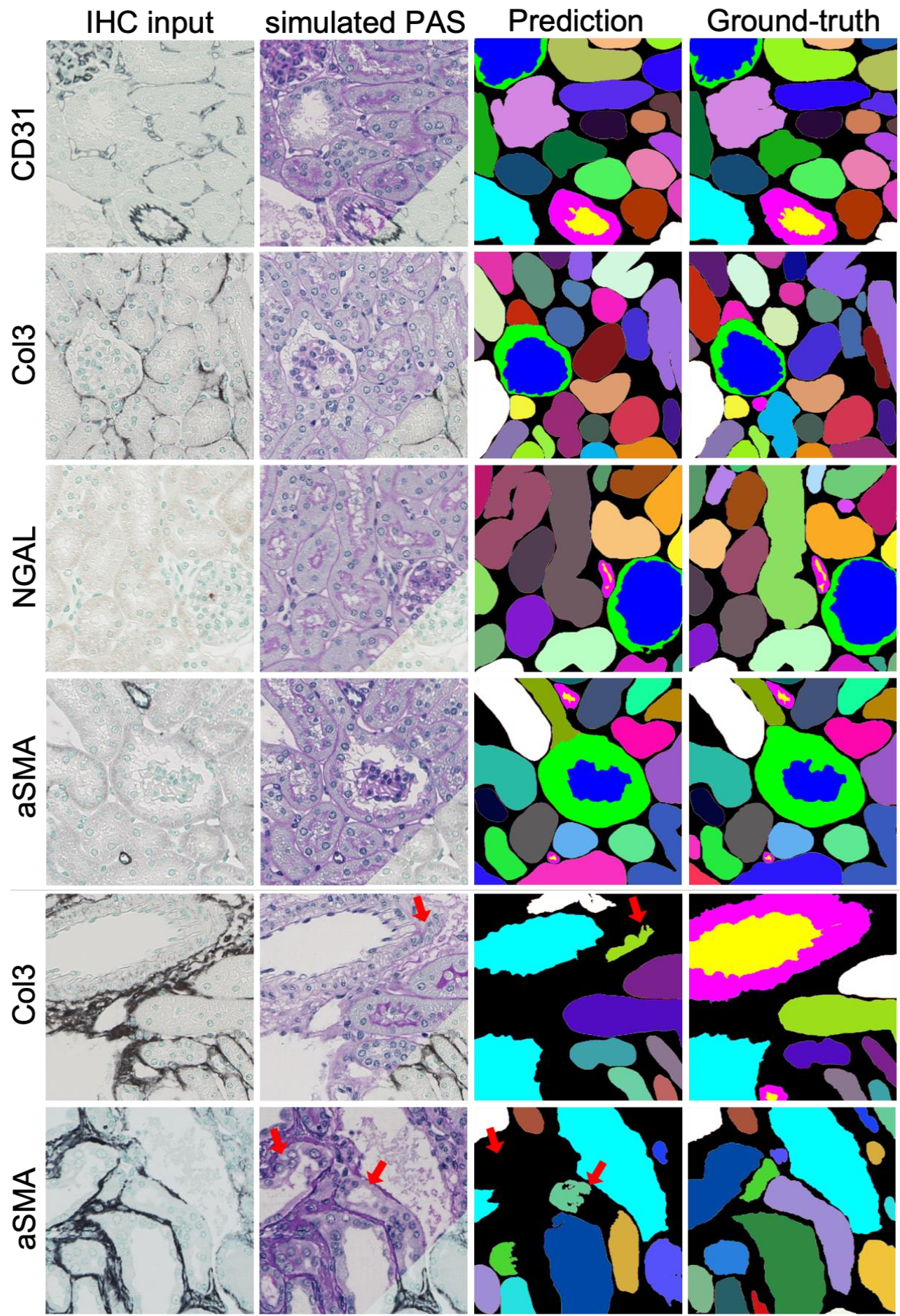

**Figure 4: Qualitative results in all stains.**

Qualitative translation and prediction results of our best performing model (CycleGAN w/ SegNet) are depicted. Predictions (column three) are performed by propagating simulated PAS translations (column two) from IHC input images (column one) through the utilized segmentation model. They are colored in accordance with Fig. 1, but in contrast tubules are colored randomly here to analyze maintaining capabilities of instance separation. We also overlaid triangular input image croppings on the PAS translations to assess spatial consistencies and pointed out severe translation artefacts by red arrows.

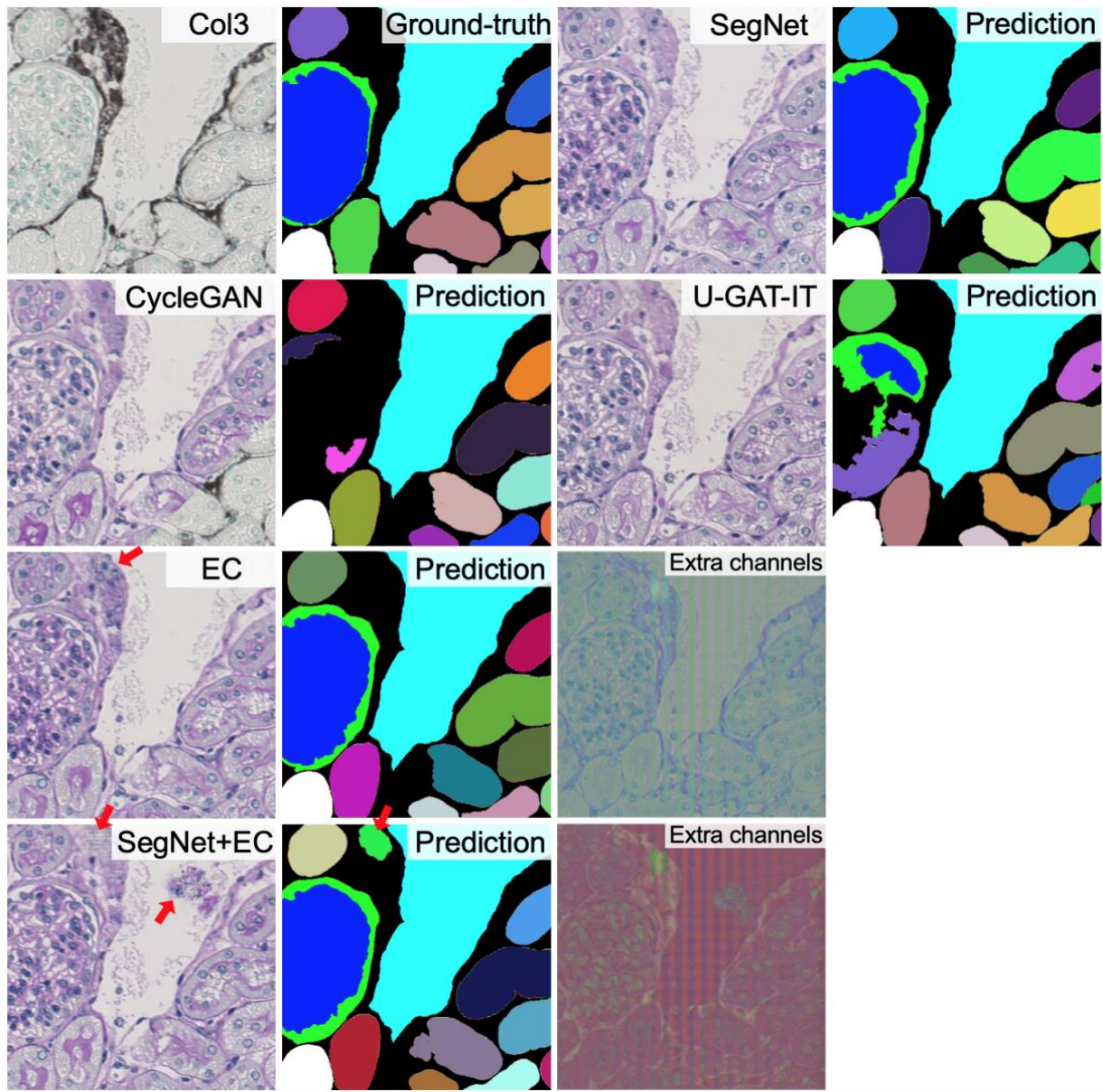

**Figure 5: Qualitative results of all translation models.**

Qualitative translation and prediction results of all translation models on an example Col3 image (upper left) are depicted. Extra channels are visualized as RGB image and distinct translation artifacts are marked by red arrows.

**Table 1. Segmentation performance quantifications on translated simulated PAS images from all IHC stains.**

Segmentation performance was measured by instance-level Dice scores and standard deviations of their underlying Dice score distributions. Those distributions were also compared in all conducted experiments against the unmodified CycleGAN in all classes using t-tests (*p<0.05 was considered statistically significant), respectively. The proposed CycleGAN variant incorporating the segmentation network and extra channels is denoted by "w/ SegNet & EC".

| **CD31** | Classes | | | | | | ∅ |
|---|---|---|---|---|---|---|---|
| | full glomerulus | glomerular tuft | tubule | artery | arterial lumen | vein | |
| CycleGAN | 87.3 ± 23.1 | 86.9 ± 21.1 | 88.6 ± 20.3 | 52.1 ± 37.9 | 56.0 ± 41.9 | 76.4 ± 38.6 | 74.5 ± 30.5 |
| w/ SegNet | **92.4 ± 15.2** | 88.9 ± 16.9 | **89.2 ± 19.9** | 53.8 ± 37.2 | **63.4 ± 40.4** | **90.4 ± 22.5*** | **79.7 ± 25.3** |
| w/ EC | 87.3 ± 25.6 | 89.7 ± 16.3 | 88.5 ± 21.2 | 50.3 ± 38.2 | 59.4 ± 42.3 | 85.5 ± 30.5 | 76.8 ± 29.0 |
| w/ SegNet & EC | 92.2 ± 17.9 | **92.3 ± 11.0*** | 87.9 ± 21.2 | **56.5 ± 36.0** | 53.7 ± 42.9 | 82.7 ± 32.7 | 77.6 ± 27.0 |
| U-GAT-IT | 79.0 ± 34.6 | 82.8 ± 26.2 | 82.8 ± 25.8* | 33.7 ± 37.0* | 43.0 ± 43.0* | 83.2 ± 32.0 | 67.4 ± 33.1 |
| **aSMA** | | | | | | | |
| CycleGAN | 73.5 ± 35.9 | 74.9 ± 31.7 | 80.6 ± 29.5 | 57.5 ± 40.7 | 55.1 ± 39.7 | 72.7 ± 38.1 | 69.1 ± 35.9 |
| w/ SegNet | **77.5 ± 35.2** | **81.5 ± 29.4** | **80.9 ± 29.4** | **69.2 ± 33.5** | **67.5 ± 34.0** | 85.9 ± 28.8 | **77.1 ± 31.7** |
| w/ EC | 72.2 ± 38.6 | 75.0 ± 36.2 | 79.2 ± 30.7 | 66.7 ± 34.0 | 55.2 ± 38.5 | **92.3 ± 7.9*** | 73.4 ± 30.9 |
| w/ SegNet & EC | 75.2 ± 37.9 | **81.5 ± 28.7** | 78.2 ± 31.4* | 62.0 ± 39.2 | 60.8 ± 36.6 | 89.1 ± 21.4 | 74.5 ± 32.5 |
| U-GAT-IT | 65.5 ± 42.0 | 70.9 ± 37.1 | 70.6 ± 34.1* | 47.7 ± 42.2 | 48.3 ± 42.1 | 79.3 ± 33.1 | 63.7 ± 38.4 |
| **Col3** | | | | | | | |
| CycleGAN | 71.0 ± 41.3 | 68.8 ± 40.2 | 84.7 ± 25.9 | 23.3 ± 34.1 | 26.1 ± 36.7 | **88.2 ± 26.8** | 60.3 ± 34.2 |
| w/ SegNet | **82.6 ± 29.5*** | **79.2 ± 28.4*** | **85.6 ± 24.8** | 28.6 ± 35.1 | **36.2 ± 39.5** | 86.7 ± 29.2 | **66.5 ± 31.1** |
| w/ EC | 72.5 ± 39.4 | 72.7 ± 35.6 | 84.8 ± 25.3 | 25.4 ± 36.1 | 24.5 ± 36.2 | 83.4 ± 31.7 | 60.6 ± 34.0 |
| w/ SegNet & EC | 75.1 ± 37.9 | 74.6 ± 34.1 | 84.9 ± 25.5 | **33.1 ± 37.0*** | 34.2 ± 37.5 | 84.5 ± 30.4 | 64.4 ± 33.7 |
| U-GAT-IT | 64.3 ± 41.2 | 64.5 ± 38.5 | 82.2 ± 27.4* | 32.2 ± 37.3 | 31.8 ± 36.9 | 85.8 ± 29.5 | 60.1 ± 35.1 |
| **NGAL** | | | | | | | |
| CycleGAN | **84.3 ± 28.4** | 85.1 ± 26.5 | 85.0 ± 26.5 | 43.7 ± 41.8 | 50.6 ± 42.0 | 85.0 ± 24.5 | 72.3 ± 31.6 |
| w/ SegNet | **84.3 ± 29.2** | 82.7 ± 29.2 | **85.6 ± 23.8** | 50.6 ± 40.7 | 55.2 ± 40.5 | 84.9 ± 30.2 | 73.9 ± 32.2 |
| w/ EC | 83.8 ± 29.5 | **85.6 ± 24.2** | 84.3 ± 24.8 | 40.8 ± 39.2 | 46.0 ± 40.5 | **89.9 ± 22.8** | 71.7 ± 30.2 |
| w/ SegNet & EC | 83.4 ± 31.1 | 82.1 ± 29.1 | 85.3 ± 23.9 | 49.8 ± 41.7 | **55.2 ± 41.9** | 71.7 ± 41.5 | 71.3 ± 34.9 |
| U-GAT-IT | 78.0 ± 36.6 | 80.7 ± 29.6 | 78.5 ± 28.0 | 33.5 ± 39.5 | 41.1 ± 42.4 | 67.6 ± 31.8 | 63.2 ± 34.7 |